\documentclass[aps,prl,twocolumn,amsmath,amssymb,showpacs,superscriptaddress,notitlepage,longbibliography]{revtex4-2}
\usepackage[colorlinks=true,linkcolor=blue,anchorcolor=red,citecolor=blue, urlcolor=blue]{hyperref}
\usepackage{bm}
\usepackage{graphicx}
\usepackage{xcolor}
\usepackage{wasysym}
\usepackage{makecell}
\newcolumntype{?}{!{\vrule width 0.5pt}}
\usepackage{multirow}

\begin{document}
\title{Laughlin charge pumping from interplay of chiral Dirac and chiral Majorana modes}

\author{Zhan Cao}
\email{caozhan@baqis.ac.cn}
\affiliation{Beijing Academy of Quantum Information Sciences, Beijing 100193, China}

\author{Yang Feng}
\affiliation{Beijing Academy of Quantum Information Sciences, Beijing 100193, China}

\author{Zhi-Hai Liu}
\affiliation{Beijing Academy of Quantum Information Sciences, Beijing 100193, China}

\author{Ke He}
\affiliation{State Key Laboratory of Low Dimensional Quantum Physics, Department of Physics, Tsinghua University, Beijing 100084, China}
\affiliation{Beijing Academy of Quantum Information Sciences, Beijing 100193, China}
\affiliation{Frontier Science Center for Quantum Information, Beijing 100084, China}
\affiliation{Hefei National Laboratory, Hefei 230088, China}

\begin{abstract}
Laughlin charge pumping has provided critical insights into the topological classification of individual materials, but remains largely unexplored in topological junctions. We explore Laughlin charge pumping in junctions composed of a chiral topological superconductor sandwiched between two quantum anomalous Hall insulators, driven by an adiabatically varying magnetic flux. Here, charge pumping can be mediated merely by chiral Dirac modes or by the interplay of chiral Dirac and chiral Majorana modes (CMMs). In the former case, a variation of one magnetic flux quantum induces the pumping of a unit charge, as the chiral Dirac mode accumulates the full flux-induced phase. In contrast, in the latter case, pumping a unit charge requires a variation of fractional magnetic flux quanta, determined by the device geometry and the parity of the number of enclosed superconducting vortices. This unique feature results from the charge-neutral and zero-momentum nature of zero-energy CMMs. Our work offers an experimentally viable pathway toward detecting CMMs and could also inspire further research into Laughlin charge or spin pumping in diverse topological junctions, which are now within experimental reach.
\end{abstract}

\maketitle
The celebrated Laughlin's charge pump~\cite{laughlin1981quantized} operates in a quantum Hall cylinder threaded by an axial magnetic flux. When the flux is adiabatically varied by one flux quantum, quantized charges are transferred through the cylinder.
Laughlin charge pumping has been previously observed in two-dimensional electron gases~\cite{dolgopolov1992quantum,jeanneret1995observation}, and more recently, in ultracold atomic gases~\cite{fabre2022laughlin} and magnetic heterostructures of topological insulators~\cite{kawamura2023laughlin}.
The number of quantized charges is linked to the topological invariant, i.e., the Chern number, of quantum Hall states~\cite{thouless1982quatized,kohmoto1985topological}. This profound connection has spurred the topological classification of band insulators through the Laughlin pumping of quantized charge or spin~\cite{fu2006time,meidan2010optimal,meidan2011topological,fulga2012scattering}. Recently, junctions composed of quantum anomalous Hall insulators (QAHIs)~\cite{qi2006topological,yu2010quantized} with different Chern numbers have been realized~\cite{ovchinnikov2022topological,zhao2023creation}, opening the possibility of integrating distinct topological phases within a single sample. This progress inspires us to explore the manifestation of Laughlin charge pumping in topological junctions.

In this Letter, we study Laughlin charge pumping in a junction composed of QAHI and chiral topological superconductor (CTSC)~\cite{qi2010topological,sato2017topological}. Both materials feature gapped bulk states and gapless edge modes---chiral Dirac modes (CDMs) in QAHI and chiral Majorana modes (CMMs) in CTSC. CDMs propagate unidirectionally and are topologically protected against nonmagnetic impurities~\cite{buttiker1988absence}, making them promising for low-dissipation electronics. Tremendous progress has been made in QAHI and CDM research~\cite{liu2016quantum,chang2023quantum} since the pioneering experiment~\cite{chang2013experimental}. CMMs—like their zero-dimensional, localized counterparts, Majorana zero modes (MZMs)~\cite{read2000paired,kitaev2001unpaired}—are considered strong candidates for implementing topological quantum computation~\cite{nayak2008non,sarma2015Majorana,lian2018topological,zhou2019non}. Their propagating nature has been leveraged in the design of quantum gate operations, potentially offering speed advantages over those based on localized MZMs~\cite{lian2018topological,zhou2019non}. However, despite intensive efforts, both MZMs and CMMs remain experimentally elusive~\cite{cao2023recent,machida2024searching,mynote1,kayyalha2020absence}. The urgent need of effective CMM detection protocols motivates the integration of CTSC into our charge pump design.

\begin{figure}[b!]
\centering
\includegraphics[width=\columnwidth]{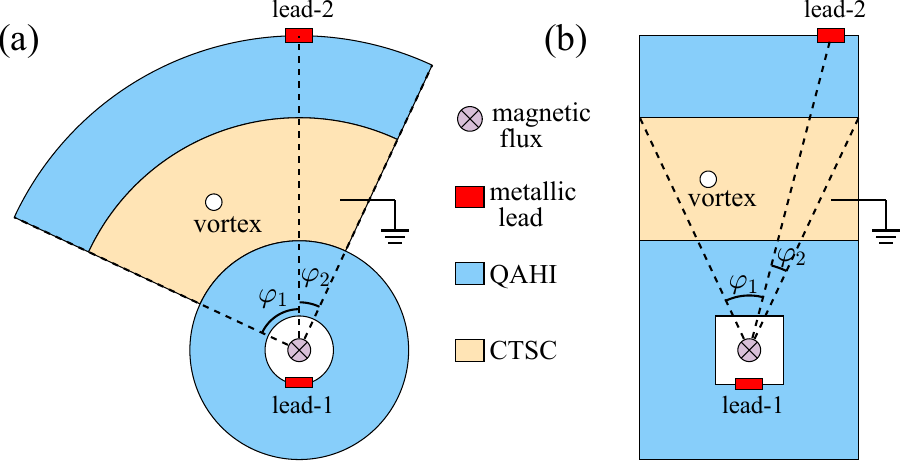}
\caption{Schematic diagrams of our charge pump based on a QAHI-CTSC-QAHI junction. Charge pumping through the leads is driven by an adiabatically varying magnetic flux confined at the center. The empty circle in the CTSC region represents a vortex. $\varphi_{1}$, $\varphi_{2}$ denote the angles between the dashed lines. Since geometries in panels (a) and (b) share the same topology, we primarily focus on panel (a) for analyses.}\label{Fig1}
\end{figure}

As schematically shown in Fig.~\ref{Fig1}, our charge pump can be designed using either of the two geometries that share the same topology. It operates by adiabatically varying the magnetic flux confined within a solenoid. We analyze the flux variation required to transfer a unit charge into or out of each lead. For lead-1, the charge pumping is mediated by a single CDM, and a variation of one flux quantum is sufficient. In contrast, for lead-2, the charge pumping arises from the interplay of CDMs and CMMs, and the required flux variation depends on multiple factors, as summarized in Table \ref{tb1}. Notably, for the $N=1$ CTSC cases, variations of fractional flux quanta are required. This is attributed to two properties of zero-energy CMMs: (1) They decouple from the magnetic vector potential due to their charge neutrality and (2) they accumulate no dynamical phase because of their zero momentum. These unique charge pumping features provide a promising avenue for detecting CMMs. In addition, no charge pumping occurs through lead-2 when a metal or conventional superconductor (SC) replaces the CTSC. Thus, our charge pump can identify whether a nominal CTSC is intact, a focus of recent experiments~\cite{kayyalha2020absence,huang2024inducing,uday2024non}.

\begin{table}[t!]
\centering
\tabcolsep=0.33cm
\renewcommand\arraystretch{1}
\caption{Required magnetic flux variation $\delta n_f\equiv \delta\Phi/\Phi_0$ for adiabatically pumping a unit charge through lead-2 in Fig.~\ref{Fig1} in three relevant cases. $\delta\Phi$ denotes variation of flux $\Phi$ and $\Phi_0= h/q$ is the flux quantum, where $h$ is the Planck constant and $q$ is the elementary charge. $N$ is the number of CMMs hosted by the CTSC and $n_v$ is the number of vortices within the CTSC. $\varphi_{\pm}\equiv\varphi_1\pm \varphi_2$, with $\varphi_{1(2)}$ the angles defined in Fig.~\ref{Fig1}. }\label{tb1}
\begin{tabular}{cccc}
\hline
\hline
&\makecell[c]{$N=2$} &\makecell[c]{$N=1$,~$n_v$=even} &\makecell[c]{$N=1$,~$n_v$=odd}\\
\hline
{$\delta n_f$}  &$1$ &$2\pi/\varphi_+$ &$2\pi/\varphi_-$\\
\hline
\hline
\end{tabular}
\end{table}

\begin{figure}[t!]
\centering
\includegraphics[width=\columnwidth]{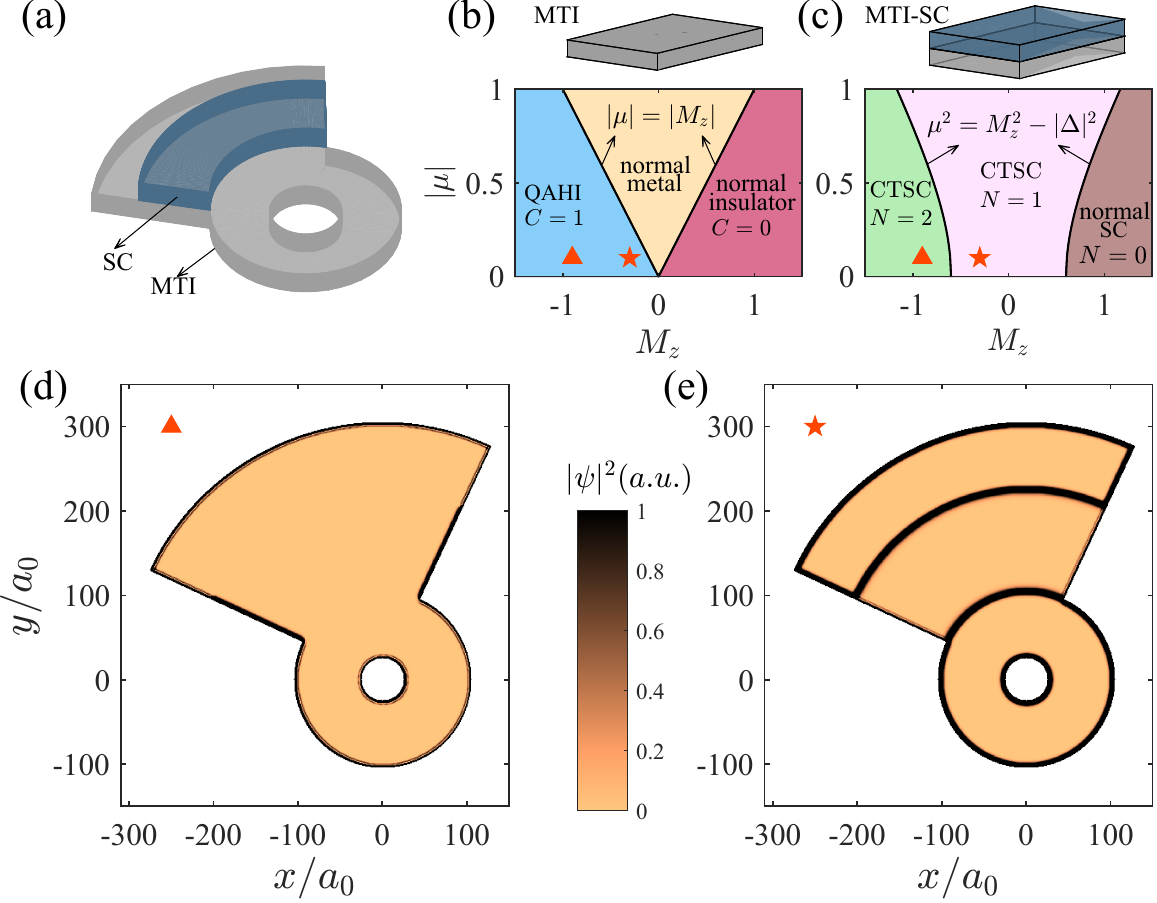}
\caption{(a) An illustration of our charge pump based on an MTI and an $s$-wave SC. (b), (c) Phase diagrams for an individual MTI and an MTI-SC hybrid, see text for details. Black lines along with expressions indicate the phase boundaries. (d), (e) Superpositions of wavefunction probability amplitudes of the 40 lowest near-zero-energy eigenstates. The parameter sets used for obtaining (d) and (e) are marked by triangles and pentagrams, respectively, in (b) and (c). $a_0$ is the lattice spacing used in simulations~\cite{supp}. }\label{Fig2}
\end{figure}

{\color{blue}\emph{Numerical simulations on a candidate platform}.} Based on the theories~\cite{wang2015chiral, wang2016electrically, chen2023lectrostatic}, our charge pump could potentially be realized by covering an $s$-wave SC on the top of a magnetic topological insulator (MTI) thin film, as illustrated in Fig.~\ref{Fig2}(a) for the geometry in Fig.~\ref{Fig1}(a). Notably, MTI–SC heterostructures matching the Fig.~\ref{Fig1}(b) geometry (without the central hole) have been fabricated~\cite{kayyalha2020absence,huang2024inducing,uday2024non}. An individual MTI and an MTI-SC hybrid can be described by the effective Hamiltonian~\cite{qi2010chiral}
\begin{eqnarray}
&&\hspace{-0.3cm}h_{\rm MTI}(  \bm k)=\left(
\begin{array}
[c]{cc}
M_z+B\bm k^{2}-\mu & D(  k_{x}-ik_{y})  \\
D(  k_{x}+ik_{y})   & -M_z-B\bm k^{2}-\mu
\end{array}
\right),\label{h_MTI}\\
&&\hspace{-0.3cm}h_{\rm MTI-SC}(\bm k)=\frac{1}{2}\left(
\begin{array}
[c]{cc}
h_{\rm MTI}(\bm k) & i\Delta\sigma_{y}\\
-i\Delta^{\ast}\sigma_{y} & -h_{\rm MTI}^{\ast}(-\bm k)
\end{array}
\right),\label{h_MTI-SC}
\end{eqnarray}
where $\bm k=(k_x,k_y)$ is the electron momentum, $\sigma_y$ is the second Pauli matrix, and $M_z$, $B$, $D$, $\mu$, and $\Delta$ are parameters representing the magnetic gap, inverse effective electron mass, Fermi velocity, chemical potential, and pairing potential, respectively. 

These two models predict rich phase diagrams~\cite{qi2010chiral}. The MTI can be a QAHI, normal metal, or normal insulator, with the insulating phases distinguished by the Chern number $C$ [Fig.~\ref{Fig2}(b)]. Similarly, the MTI-SC hybrid can be either a CTSC or a normal SC, classified by the Bogoliubov-de Gennes Chern number $N$ [Fig.~\ref{Fig2}(c)]. $C$ and $N$ also determine the numbers of CDMs and CMMs at the edges of MTI and MTI-SC hybrid, respectively. Notably, the $C=1$ QAHI is topologically equivalent (inequivalent) to the $N=2$ ($N=1$) CTSC, as a CDM can be viewed as the combination of two CMMs~\cite{qi2010chiral}. Experimentally, these phases can be accessed by tuning $\mu$ and $M_z$ using electrostatic gates and an out-of-plane magnetic field, respectively. The latter may induce vortices within the CTSC, with each vortex trapping an integer multiple of the superconducting flux quantum $\Phi_{0}^s=h/2q$~\cite{tinkham2004introduction}. In the $N=1$ CTSC, a vortex trapping an odd multiple of $\Phi_0^s$ binds a single MZM at its core~\cite{alicea2012new}. Hereafter, the term `vortices' implicitly refers to such vortices.

Our charge pump is modeled by Eqs.~\eqref{h_MTI} and \eqref{h_MTI-SC}, incorporating magnetic flux, vortices, and leads; see Sec.~SI of \cite{supp} for detailed descriptions of the model and numerical simulations. To simulate charge pumps composed of the $C=1$ QAHI and the CTSC with $N=2$ or $N=1$, we use two representative parameter sets marked in Figs.~\ref{Fig2}(b) and \ref{Fig2}(c). In the absence of leads and magnetic flux, we calculate the wave functions of the 40 lowest near-zero-energy eigenstates, whose probability amplitudes are superposed in Figs.~\ref{Fig2}(d) and \ref{Fig2}(e). The dark regions highlight the spatial distributions of topological edge modes. Since they are expected to arise at interfaces between topologically inequivalent gapped phases, we identify CDMs at the QAHI-vacuum interfaces and CMMs at the CTSC-vacuum interfaces in both Figs.~\ref{Fig2}(d) and \ref{Fig2}(e), while CMMs appear at the QAHI-CTSC interfaces only in Fig.~\ref{Fig2}(e). 

We define $n_f(t)\equiv \Phi(t)/\Phi_0$ to measure the applied flux $\Phi(t)$. At time $t$, the cumulative pumped charge (in units of the elementary charge $q$) through lead-$m$ ($m=\{1,2\}$) in the zero-temperature limit can be evaluated using the adiabatic scattering formulation~\cite{meidan2011topological,brouwer1998scattering,blaauboer2002charge,taddei2004andreev,alos2014adiabatic,luo2021half}
\begin{equation}
Q_{m}(t)=\sum_{m^\prime \eta}\int_{0}^{t}\frac{d\tau}{2\pi}\sigma_\eta\operatorname{Im}\textrm{Tr}\bigg[\frac{d S_{m'm}^{\eta e}}{d n_f}S_{m'm}^{\eta e\dagger}\bigg]\frac{d n_f}{d \tau},\label{Qmte}
\end{equation}
where $\eta=\{e,h\}$, $\sigma_{e(h)}=\pm 1$, and $S_{m'm}^{\eta e}$ is the scattering matrix describing the conversion of an electron at the Fermi level (set to zero) in lead-$m$ into an electron ($\eta=e$) or hole ($\eta=h$) in lead-$m'$. The scattering matrices can be obtained by the Mahaux-Weidenm\"{u}ller formula~\cite{mahaux1969shell}, as elaborated in Sec.~SI of \cite{supp}. We focus on the case where $n_f$ increases linearly in time and numerically compute the $S_{m'm}^{\eta e}$ at zero-energy, see Sec.~SI of \cite{supp} for details. 

\begin{figure}[t!]
\centering
\includegraphics[width=\columnwidth]{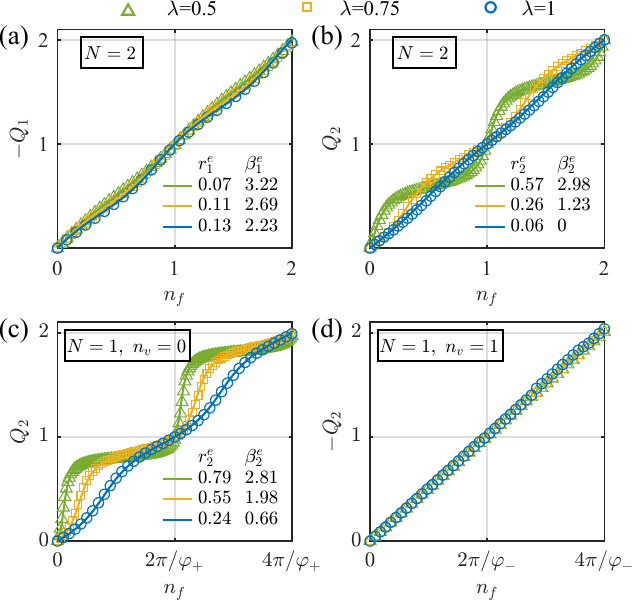}
\caption{Simulations (markers) and fittings (solid lines) of cumulative pumped charge through (a) lead-1 and (b)--(d) lead-2 for the cases indicated in the boxes. $\lambda$ is the electron hopping amplitude at the lead-QAHI interfaces. Fittings in (a)--(c) are based on Eq.~\eqref{Q1} and the $Q_2$ expressions in Table \ref{tb2}, with fitting parameters $r_m^e$ and $\beta_m^e$. The linear behavior in (d) agrees with the analytical $Q_2$ in Table \ref{tb2}. Parameters used in the simulations are the same as Figs.~\ref{Fig2}(d) and \ref{Fig2}(e).}\label{Fig3}
\end{figure}

To simulate realistic charge pumping in the presence of leads, we examine various electron hopping amplitudes $\lambda$ at the lead-QAHI interfaces. The markers in Fig.~\ref{Fig3} show the simulated results of $Q_1$ and $Q_2$ in different CTSC cases. We are interested in the flux variation $\delta n_f$ required to vary $Q_1$ or $Q_2$ by one unit. Figure \ref{Fig3}(a) shows that $\delta n_f=1$ for $Q_1$. This result is tied to the inner QAHI edge and thus is independent of the properties of the CTSC. In contrast, Figs.~\ref{Fig3}(b)--\ref{Fig3}(d) show that $\delta n_f$ for $Q_2$ depends on multiple factors: the geometry parameters $\varphi_{\pm}\equiv\varphi_1\pm \varphi_2$ (see $\varphi_{1,2}$ in Fig.~\ref{Fig1}), the Chern number $N$ of the CTSC, and the number $n_v$ of vortices within the CTSC. Specifically, $\delta n_f=1$ [Fig.~\ref{Fig3}(b)] for the $N=2$ CTSC, while $\delta n_f=2\pi/\varphi_+$ [Fig.~\ref{Fig3}(c)] or $\delta n_f=2\pi/\varphi_-$ [Fig.~\ref{Fig3}(d)] for the $N=1$ CTSC cases. Moreover, in Fig.~\ref{Fig3}(d), $Q_2$ varies linearly with $n_f$, independent of the value of $\lambda$. These unique charge pumping features also appear in the pump geometry in Fig.~\ref{Fig1}(b), as demonstrated in Sec.~SII of \cite{supp}. To facilitate identifying fractional flux variations, we suggest fabricating devices in which $\varphi_+$ and $\varphi_-$ deviate significantly from $2\pi$ and 0, respectively.

\begin{figure}[t!]
\centering
\includegraphics[width=\columnwidth]{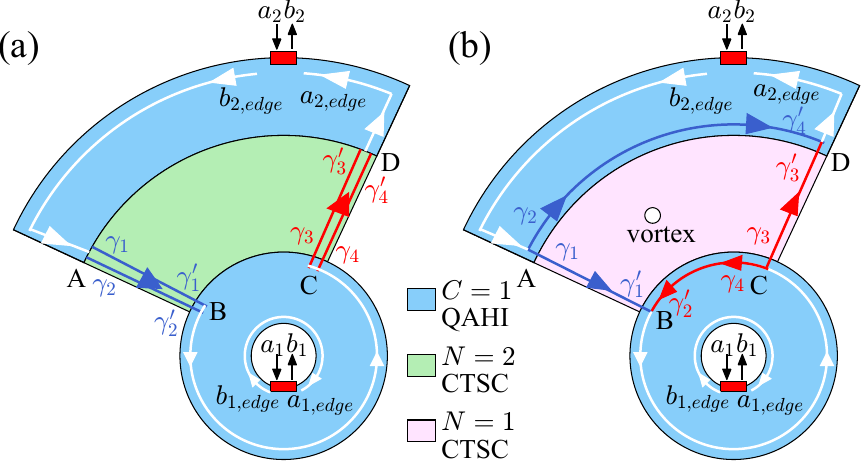}
\caption{Propagation routes of topological edge modes. While the QAHI edges host one CDM (white lines), the CTSC edges can host (a) two or (b) one CMM (red and blue lines). CDMs convert into CMMs at vertices $A$ and $C$, while inverse processes occur at vertices $B$ and $D$.
}\label{Fig4}
\end{figure}

{\color{blue}\emph{Mechanisms of the unique charge pumping}.} To unveil the underlying pumping mechanisms, we perform model-independent analytical analyses with reasonable assumptions: (1) To derive the zero-energy scattering matrices in Eq.~\eqref{Qmte}, we consider only the gapless edge modes of the pump and neglect bulk contributions---an approximation valid for systems with sufficiently large sizes and considerable topological gaps, as in Figs.~\ref{Fig2}(d) and \ref{Fig2}(e). (2) Since a CDM effectively couples to a single conduction channel in a lead, we simplify the scattering matrices in Eq.~\eqref{Qmte} to $b_{m}^{\eta}=\sum_{ m^\prime \eta^\prime}S_{mm^\prime}^{\eta\eta^\prime}a_{m^\prime}^{\eta^\prime}$, where $a^\eta_m$ ($b^\eta_m$) denotes the amplitude of the incoming (outgoing) $\eta$-type single-mode in lead-$m$. (3) As the inner and outer edge modes in Figs.~\ref{Fig2}(d) and \ref{Fig2}(e) are well-separated, we neglect the charge transfer between lead-1 and lead-2.

Based on our earlier analysis of Figs.~\ref{Fig2}(d) and \ref{Fig2}(e), we depict the spatial distributions of CDMs (white lines) and CMMs (red and blue lines) in Fig.~\ref{Fig4}. The arrows indicate the chirality of the edge modes. We adopt the convention that electron-type CDMs propagating clockwise accumulate negative flux-induced phases. At the interface between lead-$m$ and its nearest QAHI edge, the amplitudes $a_{m,edge}^\eta$ and $b_{m,edge}^\eta$ of incoming and outgoing $\eta$-type CDMs are formally related to the amplitudes $a^\eta_{m}$ and $b^\eta_{m}$ by a scattering matrix~\cite{moskalets2011scattering}
\begin{equation}
\left(
\begin{array}
[c]{c}%
b^\eta_{m}\\
b^\eta_{m,edge}
\end{array}
\right)  =e^{i\alpha_m^\eta}\left(
\begin{array}
[c]{cc}%
r^\eta_m & it^\eta_m\\
it^\eta_m & r^\eta_m
\end{array}
\right)  \left(
\begin{array}
[c]{c}%
a^\eta_{m}\\
a^\eta_{m,edge}
\end{array}
\right),\label{sm_junction}
\end{equation}
where the real numbers $r_m^\eta$ and $t_m^\eta=\sqrt{1-(r_m^\eta)^2}$ are reflection and transmission amplitudes at both sides of the interface, and $\alpha_m^\eta$ is associated with the effective charge at the interface~\cite{moskalets2011scattering}. The intrinsic particle-hole symmetry imposes $r_{m}^h=r_{m}^{e}$, $t_{m}^h=-t_{m}^{e}$, and $\alpha_m^h=-\alpha_m^e$.

We first derive $Q_1$. In Figs.~\ref{Fig4}(a) and \ref{Fig4}(b), the circular motion of an electron-type CDM along the inner QAHI edge accumulates a flux-induced phase $e^{- 2i\pi n_f}$ and a dynamical phase $e^{- i\theta_d}$. Due to spin-momentum locking~\cite{ando2013topological}, this motion is accompanied by a $2\pi$ spin rotation, giving rise to a $\pi$ Berry phase~\cite{sakurai2020modern}. Therefore, $a^{e}_{1,edge}=e^{- i(2\pi n_f+\theta_d+\pi)}b^{e}_{1,edge}$. Combining this with Eq.~\eqref{sm_junction} yields
\begin{eqnarray}
&&S_{11}^{ee}=H(-n_f,r_1^e,\alpha_1^e,\beta_1^e),~~S_{11}^{he}=0, \label{S1}\\
&&H(n_f,r,\alpha,\beta)=\frac{r+e^{i(2\pi n_f+\beta)}}{1+re^{i(2\pi n_f+\beta)}}e^{i\alpha},\label{Htr}
\end{eqnarray}
where $\beta_1^e=\alpha_1^e-\theta_d$. For a linearly increasing $n_f$ over time, substituting Eq.~\eqref{S1} into Eq.~\eqref{Qmte} gives (see Sec.~SIV of \cite{supp})
\begin{eqnarray}
Q_{1}&=&F(-n_f,r_1^e,\beta_1^e),\label{Q1}\\
F(n_f,r,\beta)&=&n_f-\frac{1}{\pi}\arg\frac{1+ re^{i(2\pi n_f+\beta)}}{1+ re^{i\beta}}.\label{Ftr}
\end{eqnarray}
Evidently, the second term of $F$ vanishes for integer $n_f$, indicating that a variation of one flux quantum pumps exactly a unit charge through lead-1. As shown in Fig.~\ref{Fig3}(a), the simulations of $Q_1$ are well fit (solid lines) by Eq.~\eqref{Q1} using parameters $r_1^e$ and $\beta_1^e$. 

We proceed to derive $Q_2$ along the same lines. As illustrated in Figs.~\ref{Fig4}(a) and \ref{Fig4}(b), three scattering processes are common: (1) The electron-type and hole-type CDMs leaving lead-2 convert into two CMMs $\gamma_{1,2}$ at vertex $A$; (2) two CMMs $\gamma^\prime_{1,2}$ convert into CDMs at vertex $B$ and then again into CMMs $\gamma_{3,4}$ at vertex $C$; (3) two CMMs $\gamma^\prime_{3,4}$ convert into CDMs toward lead-2. These scattering processes can be formulated as
\begin{eqnarray}
&&\left(
\begin{array}
[c]{c}
\gamma_{1}\\
\gamma_{2}
\end{array}
\right)=S_{A}^{in}\left(
\begin{array}
[c]{cc}
e^{i\chi_1} & 0\\
0 &e^{-i\chi_1}
\end{array}
\right)  \left(
\begin{array}
[c]{c}
b_{2,edge}^e\\
b_{2,edge}^{h}
\end{array}
\right),\label{SM1}\\
&&\left(
\begin{array}
[c]{c}
\gamma_{3}\\
\gamma_{4}
\end{array}
\right)=S_{C}^{in}\left(
\begin{array}
[c]{cc}
e^{i\chi_3} & 0\\
0 & e^{-i\chi_3}
\end{array}
\right)  S_{B}^{out}\left(
\begin{array}
[c]{c}
\gamma^\prime_{1}\\
\gamma^\prime_{2}
\end{array}
\right),\label{SM3}\\
&&\left(
\begin{array}
[c]{c}
a_{2,edge}^e\\
a_{2,edge}^{h}
\end{array}
\right)=\left(
\begin{array}
[c]{cc}
e^{i\chi_2} & 0 \\
0 &e^{-i\chi_2}
\end{array}
\right)  S_{D}^{out}\left(
\begin{array}
[c]{c}
\gamma^\prime_{3}\\
\gamma^\prime_{4}
\end{array}
\right),\label{SM2}
\end{eqnarray}
where $\chi_1=\varphi_1 n_f+\theta_{d1}$, $\chi_2=\varphi_2 n_f+\theta_{d2}$, and $\chi_3=(2\pi-\varphi_1-\varphi_2)n_f+\theta_{d3}+\pi$. They include both flux-induced ($n_f$-terms) and dynamical ($\theta$-terms) phases picked up by the CDMs. $\chi_3$ also absorbs the $\pi$ Berry phase. The particle-hole symmetry and the unitarity of scattering matrices follow $S_X^{in}=\frac{1}{\sqrt{2}}\left(
\begin{array}
[c]{cc}
1 & 1\\
i & -i
\end{array}
\right)$ and $S_X^{out}=S_X^{in\dagger}$~\cite{fu2009probing,akhmerov2009electrically}.

The edge modes around the CTSC are quite distinct in Figs.~\ref{Fig4}(a) and \ref{Fig4}(b), leading to different relations between the amplitudes $\gamma$ and $\gamma^\prime$ appearing in Eqs.~\eqref{SM1}--\eqref{SM2}. In Fig.~\ref{Fig4}(a), the scattering between two CMMs along the same edge of the $N=2$ CTSC is effectively described by planar rotations~\cite{li2012scattering,li2019majorana}:
$\left(
\begin{array}
[c]{c}%
\gamma^{\prime}_{1}\\
\gamma^{\prime}_{2}
\end{array}
\right)  =U(\delta_1)  \left(
\begin{array}
[c]{c}
\gamma_{1}\\
\gamma_{2}
\end{array}
\right)$ and $\left(
\begin{array}
[c]{c}
\gamma^{\prime}_{3}\\
\gamma^{\prime}_{4}
\end{array}
\right)  =U(\delta_2) \left(
\begin{array}
[c]{c}
\gamma_{3}\\
\gamma_{4}
\end{array}
\right)$, with $U(\delta)  =\left(
\begin{array}
[c]{cc}
\cos\delta & \sin\delta\\
-\sin\delta & \cos\delta
\end{array}
\right)$. 
Substituting these relations into Eqs.~\eqref{SM1}--\eqref{SM2} yields 
\begin{equation}
a_{2,edge}^{e}=e^{ i(2\pi n_f+\phi_{0}+\pi)}b_{2,edge}^{e},\label{edge2}
\end{equation}
with $\phi_0=\theta_{d1}+\theta_{d2}+\theta_{d3}+\delta_{1}+\delta_{2}$. The edge modes accumulate the full flux-induced phase $2\pi n_f$, since the two CMMs between vertices $A$ and $B$ ($C$ and $D$) in Fig.~\ref{Fig4}(a) are equivalent to a CDM, as mentioned earlier.

In Fig.~\ref{Fig4}(b), the amplitudes associated with the CMMs obey the relations $\gamma_1^\prime=\gamma_1e^{i\delta_{AB}}$, $\gamma_2^\prime=\gamma_4e^{i\delta_{BC}}$, $\gamma_3^\prime=\gamma_3e^{i\delta_{CD}}$, and $\gamma_4^\prime=\gamma_2e^{i(\delta_{AD}+\pi+n_v\pi)}$, with $\delta_{XY}$ the dynamical phase accumulated by a CMM as it propagates between vertices $X$ and $Y$. The last relation involves the $\pi$ Berry phase and the phase windings~\cite{ivanov2001nonabelian,alicea2012new,sato2016majorana,note2} associated with the vortices, each of which is assumed to bind a single MZM. Importantly, these relations do not include any flux-induced phases. This is because charge-neutral CMMs decouple from magnetic vector potential, as discussed in Sec.~SIII of \cite{supp}. Substituting these relations into Eqs.~\eqref{SM1}--\eqref{SM2} yields 
\begin{equation}
\left(
\begin{array}
[c]{c}
a_{2,edge}^{e}\\
a_{2,edge}^{h}
\end{array}
\right)  =\left(
\begin{array}
[c]{cc}
e^{ip_+  }g_{+} & e^{-ip_-  }g_{-}\\
e^{ip_-  }g_{-} & e^{-ip_+  }g_{+}
\end{array}
\right)  \left(
\begin{array}
[c]{c}
b_{2,edge}^{e}\\
b_{2,edge}^{h}
\end{array}
\right),\label{SM7}
\end{equation}
where $p_{\pm}=\chi_1\pm \chi_2$ and $g_{\pm}=\pm \frac{1}{2}e^{i(\delta_{AD}+\pi+n_v\pi)}+\frac{\cos \chi_{3}-e^{i\delta_{BC}}}{2(1-\cos \chi_{3}e^{i\delta_{BC}})}e^{i(  \delta_{AB}+\delta_{CD})  }$. For the $N=1$ CTSC, the hosted CMM with velocity $v_M$ exhibits the dispersion $E=\hbar v_Mk$~\cite{sato2016majorana,park2014absence,supp}. Therefore, zero-energy CMMs have zero momenta, so $\delta_{AB}=\delta_{BC}=\delta_{CD}=\delta_{AD}=0$, which reduces Eq.~\eqref{SM7} to
\begin{eqnarray}
&&a_{2,edge}^{e}=e^{i(\varphi_+n_f +\theta_++\pi)}b_{2,edge}^{e},\label{edge3}\\
&&a_{2,edge}^{e(h)}=e^{\mp i(\varphi_{-}n_f+\theta_{-}\mp\pi)  }b_{2,edge}^{h(e)},\label{edge4}
\end{eqnarray}
for $n_v$ being an even or odd number, respectively, with $\varphi_\pm=\varphi_1\pm\varphi_2$ and $\theta_\pm=\theta_{d1}\pm\theta_{d2}$. Remarkably, the prefactors $\varphi_\pm$ of $n_f$ indicate that the edge modes do not accumulate the full flux-induced phase. As is evident from the above derivations, this results from the charge-neutral and zero-momentum nature of zero-energy CMMs.

\begin{table}[t!]
\centering
\tabcolsep=0.05cm
\renewcommand\arraystretch{1.2}
\caption{Analytical results related to lead-2 for different CTSC cases. $\beta_{2a}^e=\alpha_2^e+\phi_0$, $\beta_{2b}^e=\alpha_2^e+\theta_+$, and the functions $H$ and $F$ are defined in Eqs.~\eqref{Htr} and \eqref{Ftr}, respectively. }\label{tb2}
\begin{tabular}{cccc}
\hline
\hline
 &{$N=2$} &$N=1$, $n_v$=even &$N=1$, $n_v$=odd\\
\hline
$S_{22}^{ee}$  &$H(n_f,r_2^e,\alpha_2^e,\beta_{2a}^e)$ &$H(\frac{\varphi_+}{2\pi}n_f,r_2^e,\alpha_2^e,\beta_{2b}^e)$ &$0$\\
\hline
$S_{22}^{he}$   &0 &0 &$-e^{i( \varphi_{-}n_f+\theta_{-})}$\\
\hline
$Q_2$  &$F\big(n_f,r^e_2,\beta^e_{2a}\big)$ &$F\big(\frac{\varphi_+}{2\pi} n_f,r^e_2,\beta^e_{2b}\big)$ &$-\frac{\varphi_-}{2\pi}n_f$\\
\hline
\hline
\end{tabular}
\end{table}

By combining Eqs.~\eqref{Qmte} and \eqref{sm_junction}, and each of Eqs.~\eqref{edge2}, \eqref{edge3}, and \eqref{edge4}, we derive the corresponding scattering matrices and cumulative pumped charge, as listed in Table \ref{tb2}, whose fourth row implies the results in Table \ref{tb1}. The simulated $Q_2$ in Figs.~\ref{Fig3}(b) and \ref{Fig3}(c) are well fit (solid lines) by the formulas in Table \ref{tb2}. The fitted values indicate that smaller reflection amplitudes $r_m^e$ result in smoother charge pumping curves. The linear behavior in Fig.~\ref{Fig3}(d) is also captured by our analytical expression. 

{\color{blue}\emph{Discussions}.} The above analyses are performed at zero temperature. At finite temperatures, the cumulative pumped charge is obtained by integrating Eq.~\eqref{Qmte} multiplied by $-\partial_E f(E)$ over $E$~\cite{taddei2004andreev,moskalets2011scattering}, where $f(E)$ is the Fermi distribution. Given the aforementioned linear dispersion, $E=\hbar v_Mk$, of the CMMs in the $N=1$ CTSC, finite-energy electrons traversing the CTSC accumulate nonzero dynamical phases $\delta_{XY}$. This prevents the reduction of Eq.~\eqref{SM7} to Eqs.~\eqref{edge3} and \eqref{edge4}. Consequently, the observation of the fractional charge pumping requires sufficiently low temperatures, where dephasing effects that disrupt phase accumulation are also suppressed.

We discuss scenarios where a metal or conventional SC replaces the CTSC. While half-quantized two-terminal conductance plateaus in QAHI-CTSC-QAHI junctions were initially considered a hallmark of CMMs~\cite{chung2011conductance}, later theories~\cite{chen2017effects,ji2018half,li2018noise} and recent experiments~\cite{kayyalha2020absence,huang2024inducing,uday2024non} suggest that such plateaus can also arise if the CTSC behaves as a metal. In our charge pump, however, replacing the CTSC with a metal prevents charge pumping through lead-2. This is because the edge modes connected to lead-2 do not form a phase-coherent loop enclosing the applied flux, due to the absence of gap-protected chiral edge modes at the metal-vacuum interfaces. The same reasoning and conclusion apply to a conventional SC, even if chiral Andreev modes form at the QAHI-SC interfaces~\cite{zhao2020interference}.

{\color{blue}\emph{Summary}.} We have unveiled unique Laughlin charge pumping in QAHI-CTSC-QAHI junctions driven by an adiabatically varying magnetic flux. Specifically, when the CTSC hosts a single CMM, pumping a unit charge through the outer lead requires a variation of fractional magnetic flux quanta, due to the interplay of CDMs and CMMs. Observing the predicted unique charge pumping features would provide strong evidence for the existence of CMMs. Similar junction architectures have been proposed as building blocks for CMM-based topological quantum computation~\cite{lian2018topological,zhou2019non}. These junctions could potentially be realized in MTI-SC hybrid systems, which are actively studied experimentally~\cite{shen2020spectroscopic,chen2022superconducting,xu2022proximity,dong2024proximity,atanov2024proximity,uday2024induced}.
Our findings also suggest that diverse topological junctions may exhibit distinctive Laughlin charge or spin pumping behaviors, with potential fundamental or practical implications for mesoscopic transport.

{\color{blue}\emph{Acknowledgements}.} We thank W. Luo for helpful discussions. This work was supported by the National Natural Science Foundation of China (Grants No.~12374158, No.~92365201, No.~12074039, and No.~12474497) and the Innovation Program for Quantum Science and Technology (Grant No.~2021ZD0302502).

{\color{blue}\emph{Data availability}.} The data that support the findings of this article are openly available~\cite{zenodo14978710}.


%

\end{document}


\title{Supplemental Material\\
for\\
``Laughlin charge pumping from interplay of chiral Dirac and chiral Majorana modes"}

\author{Zhan Cao}
\email{caozhan@baqis.ac.cn}
\affiliation{Beijing Academy of Quantum Information Sciences, Beijing 100193, China}

\author{Yang Feng}
\affiliation{Beijing Academy of Quantum Information Sciences, Beijing 100193, China}

\author{Zhi-Hai Liu}
\affiliation{Beijing Academy of Quantum Information Sciences, Beijing 100193, China}

\author{Ke He}
\affiliation{State Key Laboratory of Low Dimensional Quantum Physics, Department of Physics, Tsinghua University, Beijing 100084, China}
\affiliation{Beijing Academy of Quantum Information Sciences, Beijing 100193, China}
\affiliation{Frontier Science Center for Quantum Information, Beijing 100084, China}
\affiliation{Hefei National Laboratory, Hefei 230088, China}


\maketitle
\tableofcontents
\section{Model Hamiltonians and simulation details of our proposed charge pump}\label{sec1}
In this work, we investigate a charge pump based on junctions of two types of topological materials, quantum anomalous Hall insulator (QAHI) and chiral topological superconductor (CTSC). We propose to realize this charge pump by a magnetic topological insulator (MTI) thin film and an $s$-wave superconductor (SC). As explained in the main text, we can employ Eqs.~(1) and (2) to model the QAHI and CTSC regions, respectively. 

The charge pump is driven by an adiabatically varying magnetic flux $\Phi$, which is confined within a solenoid threading the center hole of the pump. To take into account the applied magnetic flux, we make the substitution $\bm k\rightarrow\bm k+e\bm A/\hbar$ with the solenoid gauge $\bm A=(A_x,A_y)$, 
\begin{equation}
A_x(x,y)=-\frac{\Phi}{2\pi}\frac{  y-y_{0} }{[  (  x-x_{0})  ^{2}+(y-y_{0})  ^{2}]  },~A_y(x,y)=\frac{\Phi}{2\pi}\frac{  x-x_{0}}{[  (  x-x_{0})  ^{2}+(y-y_{0})  ^{2}]  },\label{gauge}
\end{equation}
where $(x_0,y_0)=(0,0)$ is coordinate of the center of the pump. 

In numerical simulations, one of the cases under consideration is the presence of a vortex inside the $N=1$ CTSC. This vortex may arise due to an out-of-plane magnetic field applied to tune the magnetic gap of the MTI, as mentioned in the main text. The magnetic flux trapped in the vortex is quantized in units of the superconducting flux quantum $h/2q$. The amount of trapped flux remains constant during variations of the magnetic flux constrained within the solenoid. We assume that the vortex traps one superconducting flux quantum, such that it binds a single Majorana zero mode at its core~\cite{alicea2012new}. To model the vortex, we replace the homogenous and real superconducting pairing potential $\Delta(x,y)=|\Delta|$ with $\Delta(x,y)=|\Delta|e^{i\theta(x,y)}$~\cite{penaranda2020even,jose2023theory}, where $\theta(x,y)=\arg ((x-x_v)+i(y-y_v))$, and $(x_v,y_v)$ is the coordinate of the vortex core. We assume that $(x_v,y_v)$ is located deep within the CTSC region to neglect the interaction between the vortex and edge modes.

We model the metallic leads using the Hamiltonian
\begin{equation}
H_\textrm{lead}=\sum_{\bm k,\sigma} (B\bm k^2-\mu_\textrm{lead}) c_{\bm k,\sigma}^\dag c_{\bm k,\sigma},\label{Hlead}
\end{equation}
where $\sigma=\{\uparrow,\downarrow\}$ denotes the spin, $B$ is the inverse effective electron mass, and $\mu_\textrm{lead}$ is the chemical potential of the leads. 

The numerical simulations are performed by discretizing the $\bm k$-space Hamiltonians on a square lattice with lattice space $a_0$. To be specific, the onsite and hopping Hamiltonians of the pump are given by
\begin{eqnarray}
\textbf{H}^\textrm{onsite}_\textrm{pump}(  \mathbf{i},\mathbf{i})&=&\left(
\begin{array}
[c]{cccc}
M_z+4w-\mu_\mathbf{i} &0  &0  &|\Delta|e^{i\theta_\mathbf{i}}\\
0 &-M_z-4w-\mu_\mathbf{i} &-|\Delta|e^{i\theta_\mathbf{i}} &0 \\
0 &-|\Delta|e^{-i\theta_\mathbf{i}} &-M_z-4w+\mu_\mathbf{i} &0 \\
|\Delta|e^{-i\theta_\mathbf{i}} &0  &0  & M_z+4w+\mu_\mathbf{i}
\end{array}
\right), 
\end{eqnarray}
\begin{eqnarray}
\textbf{H}^\textrm{hop}_\textrm{pump}(  \mathbf{i}+a_0\hat{x},\mathbf{i})  =\left(
\begin{array}
[c]{cccc}
-we^{i\phi_x} & -ive^{i\phi_x} &0  &0 \\
-ive^{i\phi_x} & we^{i\phi_x} &0  &0 \\
0 &0  & we^{-i\phi_x} & -ive^{-i\phi_x}\\
0 &0  & -ive^{-i\phi_x} & -we^{-i\phi_x}
\end{array}
\right), \textbf{H}^\textrm{hop}_\textrm{pump}(\mathbf{i},\mathbf{i}+\hat{x})=\textbf{H}^{\textrm{hop},\dag}_\textrm{pump}(  \mathbf{i}+\hat{x},\mathbf{i}),
\end{eqnarray}
\begin{eqnarray}
\textbf{H}^\textrm{hop}_\textrm{pump}(  \mathbf{i}+a_0\hat{y},\mathbf{i})  =\left(
\begin{array}
[c]{cccc}%
-we^{i\phi_y} & -ve^{i\phi_y} &0  &0 \\
ve^{i\phi_y} & we^{i\phi_y} &0  &0 \\
0 &0  & we^{-i\phi_y} & ve^{-i\phi_y}\\
0 &0  & -ve^{-i\phi_y} & -we^{-i\phi_y}
\end{array}
\right), \textbf{H}^\textrm{hop}_\textrm{pump}(\mathbf{i},\mathbf{i}+\hat{y})=\textbf{H}^{\textrm{hop},\dag}_\textrm{pump}(  \mathbf{i}+\hat{y},\mathbf{i}),
\end{eqnarray}
where $\mathbf{i}=(  i_{x},i_{y})  $, $\hat{x}$ ($\hat{y}$) is the unit vector along the $x$ ($y$) direction, $w=B/a_0^{2}$, $v=D/2a_0$, and the magnetic vector potential is introduced by Peierls substitution as the phase factors $\phi_x=\frac{1}{2}[A_x(i_x,i_y)+A_x(i_x+a_0,i_y)]a_0$, $\phi_y=\frac{1}{2}[A_y(i_x,i_y)+A_y(i_x,i_y+a_0)]a_0$. Similarly, the lattice Hamiltonian of the metallic leads read,
\begin{eqnarray}
\textbf{H}^\textrm{onsite}_\text{lead}(  \mathbf{i},\mathbf{i})&=&\left(
\begin{array}
[c]{cccc}
4w-\mu_\textrm{lead} &0  &0  &0\\
0 &4w-\mu_\textrm{lead} &0 &0 \\
0 &0 &-(4w-\mu_\textrm{lead}) &0 \\
0 &0  &0  &-(4w-\mu_\textrm{lead})
\end{array}
\right),
\end{eqnarray}
\begin{eqnarray}
\textbf{H}^\textrm{hop}_\textrm{lead}(  \mathbf{i}+a_0\hat{x},\mathbf{i}) =\textbf{H}^\textrm{hop}_\textrm{lead}(  \mathbf{i}+a_0\hat{y},\mathbf{i}) 
=\textbf{H}^\textrm{hop}_\textrm{lead}(\mathbf{i},\mathbf{i}+a_0\hat{x}) =\textbf{H}^\textrm{hop}_\textrm{lead}(\mathbf{i},\mathbf{i}+a_0\hat{y}) =\left(
\begin{array}
[c]{cccc}
-w &0 &0  &0 \\
0 &-w &0  &0 \\
0 &0  & w &0\\
0 &0  &0 &w
\end{array}
\right).
\end{eqnarray}
The hopping Hamiltonian at the lead-QAHI interface is
\begin{eqnarray}
\textbf{H}^\textrm{hop}_\textrm{lead-QAHI}=\left(
\begin{array}
[c]{cccc}
-\lambda &0 &0  &0 \\
0 &-\lambda &0  &0 \\
0 &0  &\lambda &0\\
0 &0  &0 &\lambda
\end{array}
\right).
\end{eqnarray}
where $\lambda$ is the electron hopping amplitude at the lead-QAHI interfaces, which reflects the transparency of the contacts between the metallic leads and their nearest QAHI edges.

In numerical simulations, we use the parameters $B = D = 1$, $\mu_{lead}=4$, $|\Delta| = 0.6$, and $a_0=1$. To simulate realistic charge pumping, we set the width of both leads to $10a_0$, and examine different $\lambda$ values as 0.5, 0.75, and 1. Note that the resulting lattice Hamiltonian is time-dependent since the magnetic flux $\Phi$ varies in time.

In the absence of the leads and magnetic flux, we can directly diagonalize the lattice Hamiltonian of the charge pump to obtain the lowest tens of eigenenergies and their corresponding wavefunctions.

In the presence of the leads and magnetic flux, we simulate the adiabatic charge pumping through the leads. Suppose lead-$m$ supports $N$ conducting modes. Let $a^\eta_{mn}$ and $b^\eta_{mn}$ denote the amplitudes of the $n$-th incoming and outgoing $\eta$-type modes in lead-$m$, respectively. The scattering matrices $S^{\eta\eta'}_{mm'}$ in Eq.~(3) are defined via the relation
\begin{equation}
b^\eta_{mn}=\sum_{\eta^\prime m^\prime n^\prime}\left[S^{\eta\eta^\prime}_{mm'}\right]_{nn'}a^{\eta^\prime}_{m'n'},
\end{equation}
which reduces to the simplified expression $b_{m}^{\eta}=\sum_{ m^\prime \eta^\prime}S_{mm^\prime}^{\eta\eta^\prime}a_{m^\prime}^{\eta^\prime}$ used in the analytical analysis in the main text, where we assume $N=1$ for physical considerations.
The full scattering matrix $\mathbf{S}(E)$ comprising all submatrices $S^{\eta\eta'}_{mm'}$ can be obtained using the Mahaux-Weidenm\"{u}ller formula~\cite{mahaux1969shell}
\begin{equation}
\textbf{S}(E)=\textbf{I}-2\pi i \textbf{W}^\dag \textbf{G}^r(E) \textbf{W}, 
\end{equation}
where $E$ is the particle energy, $\textbf{G}^r(E)=(E-\textbf{H}_\textrm{pump}+i \pi  \textbf{W}\textbf{W}^\dag)^{-1}$ is the retarded Green's function of the pump, and $\textbf{W}$ encodes the coupling between the pump and the leads. The zero-energy scattering matrices $S^{\eta\eta'}_{mm'}$ appearing in Eq.~(3) correspond to submatrices of $\mathbf{S}(0)$, selected by the lead indices $m, m'$ and particle-type indices $\eta, \eta'$.

We focus on the case where $n_f$ increases linearly in time, i.e., $\Phi=\Phi_0 t/T_p$, where $\Phi_0$ is the magnetic flux quantum and $T_p$ is the period over which $\Phi$ is increased by $\Phi_0$. In the adiabatic pumping regime, the concrete value of $T_p$ is irrelevant as long as it is sufficiently larger than $\hbar/E_g$ with $E_g\equiv\min\{E_g^\textrm{QAHI},E_g^\textrm{CTSC}\}$, where $E_g^\textrm{QAHI}$ and $E_g^\textrm{CTSC}$ are the insulating and superconducting gaps of the QAHI and CTSC, respectively. For any instantaneous flux, we compute the zero-energy scattering matrices appearing in Eq.~(3) in the main text using the transport package Kwant~\cite{groth2014kwant}, with the instantaneous lattice Hamiltonian of the whole system as input.

\begin{figure}[htbp]
\centering
\includegraphics[width=0.8\columnwidth]{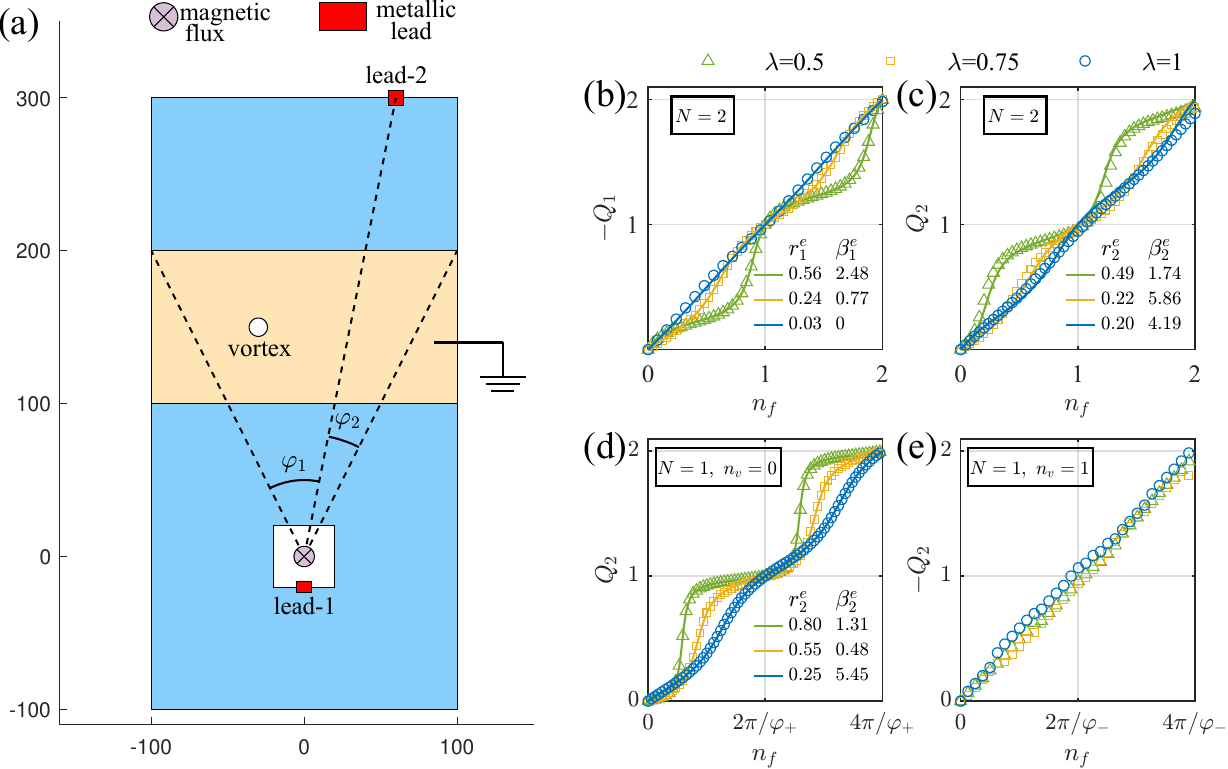}
\caption{(a) Schematic of our charge pump based on a rectangular QAHI-CTSC-QAHI junction. It has the same topology as the annular-like geometry studied in the main text. [(b)--(e)] Counterparts of Fig.~3 in the main text, but for the pump geometry shown in (a).}\label{FigS1}
\end{figure}
 
\section{Simulation results of our charge pump designed with a rectangular geometry}\label{sec2}
In the main text, we focus on the geometry shown in Fig.~1(a) for analyses. Here, we present the numerical simulations of the pump geometry shown in Fig.~1(b), which is also shown in Fig.~\ref{FigS1}(a). It is noteworthy that such a rectangular geometry has been experimentally realized~\cite{kayyalha2020absence,huang2024inducing,uday2024non}, though it lacks a central hole. For this geometry, we simulate the cumulative pumped charge through lead-1 and lead-2 in the three relevant cases considered in the main text, as indicated by the markers in Figs.~\ref{FigS1}(b)--\ref{FigS1}(e). These numerical results exhibit all the charge pumping features presented in Fig.~3, and they can also be fitted (solid lines) by the analytical formulas provided in Table II of the main text, which, in turn, confirm the charge pumping mechanisms we unveiled.

\section{Decouple of a single chiral Majorana mode from a magnetic vector potential}\label{sec3}
A chiral Dirac mode can be effectively described by the Hamiltonian~\cite{qi2010chiral}
\begin{equation}
H_\textrm{edge}=\sum_{k}(\hbar v_Fk-\mu)c_{k}^{\dag}c_{k},\label{Hedge}
\end{equation}
where $v_F$ denotes the Fermi velocity, $\mu$ is the chemical potential, and $c_{k}$ ($c_{k}^\dag$) denotes the annihilation (creation) operator of the Dirac fermion with momentum $k$. These operators satisfy the anticommunication relation $\{  c_{k}^{\dag},c_{k^{\prime}}\}  =\delta_{kk^{\prime}}$. When a magnetic vector potential $\bm A$ is present, Eq.~\eqref{Hedge} is modified as
\begin{equation}
H_\textrm{edge}=\sum_{k}\bigg[\hbar v_F\big(k+\frac{e}{\hbar}\bm A\big)-\mu\bigg]c_{k}^{\dag}c_{k}.\label{HedgeA}
\end{equation} 

Mathematically, the fermionic operator $c_k$ can be decomposed into its real and imaginary parts~\cite{qi2010chiral}: 
\begin{equation}
c_{k}=\frac{\gamma_{1,k}+i\gamma_{2,k}}{2},
\end{equation}
where $\gamma_{a,k}$ are chiral Majorana mode (CMM) operators satisfying $\gamma_{a,k}^{\dag}=\gamma_{a,-k}$ and $\{\gamma_{a,-k},\gamma_{b,k^{\prime}}\}=2\delta_{ab}\delta_{kk^{\prime}}$.  
Substituting this decomposition into Eq.~\eqref{HedgeA} gives
\begin{equation}
H_\textrm{edge}=\frac{1}{2}\sum_{k\geq0}\Big[\hbar v_Fk(\gamma_{1,-k}\gamma_{1,k}+\gamma_{2,-k}\gamma_{2,k})
+i(v_F e\bm A-\mu)(\gamma_{1,-k}\gamma_{2,k}-\gamma_{2,-k}\gamma_{1,k})\Big].
\end{equation}
It follows that the magnetic vector potential $\bm A$ and the chemical potential $\mu$ couple the two CMMs. This means that if there exist a single CMM, it is decoupled from $\bm A$ and $\mu$ must equal zero. This observation is consistent with the expectation that a charge-neutral particle cannot be affected by a magnetic vector potential, and it further justifies the linear dispersion $E=\hbar v_M k$ employed in deriving Eqs.~(14) and (15) in the main text.

\section{Derivation of Eq.~(7)}\label{sec4}
As noted in the main text, our analytical analysis neglects charge transfer between lead-1 and lead-2, i.e., we set $m^\prime=m$ in Eq.~(3). Substituting the scattering matrix obtained in Eq.~(5) into Eq.~(3), we obtain
\begin{equation}
Q_{1}(  t)  =\int_{0}^{t}\frac{d\tau}{2\pi}\operatorname{Im}\textrm{Tr}\left[  \frac{dS_{11}^{ee}}{dn_{f}}S_{11}^{ee\dag}\right]  \frac{dn_{f}(\tau)
}{d\tau}=-\int_{0}^{t}d\tau\frac{dn_{f}(\tau)}{d\tau}\frac{1-r_{1}^{e2}}{\left\vert1+r_{1}^{e}e^{-i(  2\pi n_{f}-\beta_{1}^{e})  }\right\vert ^{2}},\label{SMQ1t}
\end{equation}
where $r_{1}^{e}$ and $\beta_{1}^{e}$ are real numbers. Focusing on the linearly increasing magnetic flux $\Phi(t)=\Phi_0t/T_p$, i.e., $n_f(t)=t/T_p$, we evaluate the integral in Eq.~\eqref{SMQ1t} as
\begin{equation}
Q_{1}(  t) =-\int_{-\beta_{1}^{e}/2\pi}^{n_{f}(  t)  -\beta_{1}^{e}/2\pi}dn_{f}\frac{1-r_{1}^{e2}}{\left\vert 1+r_{1}^{e}e^{-i2\pi n_{f}}\right\vert ^{2}}=\Xi\big(n_{f}(  t)  -\beta_{1}^{e}/2\pi\big)  -\Xi\big(-\beta_{1}^{e}/2\pi\big),\label{SMQ1t2}
\end{equation}
where 
\begin{equation}
\Xi(  x)=\frac{i}{2\pi}\left[  \ln(  r_{1}^{e}+e^{i2\pi x})  -\ln (  1+r_{1}^{e}e^{i2\pi x} )  \right]=\frac{i}{2\pi}\left[  \ln (  1+r_{1}^{e}e^{-i2\pi x} )-\ln (  1+r_{1}^{e}e^{i2\pi x} )  \right]  -x =-\frac{1}{\pi}\arg (  1+r_{1}^{e}e^{-i2\pi x} )  -x.\label{SMXi}
\end{equation}
Substituting Eq.~\eqref{SMXi} into Eq.~\eqref{SMQ1t2}, we arrive at
\begin{equation}
Q_{1}(  t) =-\frac{1}{\pi}\arg\left(  1+r_{1}^{e}e^{-i\left[2\pi n_{f}(  t)  -\beta_{1}^{e}\right]  }\right)  -n_{f}(
t)  +\frac{1}{\pi}\arg\left(  1+r_{1}^{e}e^{i\beta_{1}^{e}}\right) =-n_{f}(  t)  -\frac{1}{\pi}\arg\frac{1+r_{1}^{e}e^{i\left[
-2\pi n_{f}(  t)  +\beta_{1}^{e}\right]  }}{1+r_{1}^{e}e^{i\beta_{1}^{e}}},
\end{equation}
which yields Eq.~(7) in the main text. The analytical expressions for $Q_2$ summarized in Table II are derived by applying the same integration technique to the corresponding scattering matrices.


%